\documentclass[%
 prl,
 amsmath,amssymb,
 reprint,%
]{revtex4-1}

\usepackage{graphicx}
\usepackage{dcolumn}
\usepackage{bm}

\begin{document}

\title{Multiple-Photon Absorption Attack on Entanglement-Based\\Quantum Key Distribution protocols}%

\author{Guillaume Adenier}
\email{adenier@rs.noda.tus.ac.jp}
\author{Masanori Ohya}%
\author{Noboru Watanabe}%
\affiliation{Tokyo University of Science, 2641 Yamazaki, Noda, Chiba 278-8510, Japan}%
\author{Irina Basieva}
\author{Andrei Yu. Khrennikov}
\affiliation{Linnaeus University, Vejdes plats 7, SE-351 95 V\"{a}xj\"{o}, Sweden}%

\begin{abstract}
In elaborating on the multiple-photon absorption attack on Ekert protocol proposed in arXiv:1011.4740, we show that it can be used in other entanglement-based protocols, in particular the BBM92 protocol. In this attack, the eavesdropper (Eve) is assumed to be in control of the source, and she sends pulses correlated in polarization (but not entangled) containing several photons at frequencies for which only multiple-photon absorptions are possible in Alice's and Bob's detectors. Whenever the photons stemming from one pulse are dispatched in such a way that the number of photons is insufficient to trigger a multiple-photon absorption in either channel, the pulse remains undetected. We show that this simple feature is enough to reproduce the type of statistics on the detected pulses that are considered as indicating a secure quantum key distribution, even though the source is actually a mixture of separable states. The violation of Bell inequalities measured by Alice and Bob increases with the order of the multiple-photon absorption that Eve can drive into their detectors, while the measured quantum bit error rate decreases as a function of the same variable. We show that the attack can be successful even in the simplest case of a two-photon absorption or three-photon absorption attack, and we discuss possible countermeasures, in particular the use of a fair sampling test.
\end{abstract}

\maketitle
\section{Introduction}
Quantum key distribution \cite{Gisin02} aims at preventing an eavesdropper (Eve) to acquire any information on a key distributed to two parties (Alice and Bob) who wish to use it to securely encrypt their communication through a public channel. In the ideal case, the secrecy of the key relies on two powerful theorems that are at the heart of Quantum Mechanics: the no-cloning theorem, which prevents an eavesdropper to clone an unknown quantum state, and Bell's theorem \cite{Bell64}, which guarantees that no local information actually exists which an eavesdropper could acquire. In actual implementations of the protocols, the imperfections of the components used to generate the key can be exploited by Eve to extract some information about it \cite{Scarani09}.

Here, we elaborate on the multiple-photon absorption attack on Ekert protocol \cite{Ekert91} proposed by Adenier {\it et al} \cite{Adenier10MPA}, and show that the attack can also be successful against other entanglement-based quantum key distribution (QKD) protocols, in particular the BBM92 protocol \cite{BBM92}. We will detail the performances of the attack against both Ekert and BBM92 protocols, as well as the possible countermeasures available to Alice and Bob.

\begin{figure}
\center
\includegraphics[width=7cm]{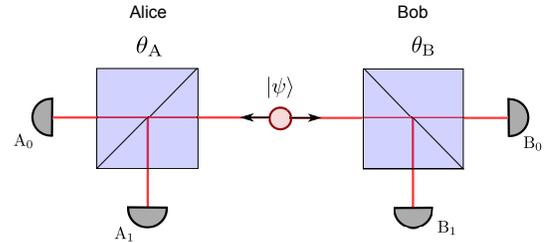}
\caption{\label{fig:protocol}Entanglement-based Quantum Cryptography. Alice and Bob randomly switch their measurement settings. In Ekert protocol, three settings are used on each sides. $\theta_\mathrm{A}=\{0,\frac{\pi}{4},\frac{\pi}{8}\}$ for Alice, and $\theta_\mathrm{B}=\{0,-\frac{\pi}{8},\frac{\pi}{8}\}$ for Bob. Detections associated with identical measurement settings ($\theta_\mathrm{A}=\theta_\mathrm{B}$) are used to produce a correlated key, while those associated to non-identical measurement settings ($\theta_\mathrm{A}\neq\theta_\mathrm{B}$) are used to check the violation of Bell inequality (and the security of the key). In the BBM92 protocol, Alice and Bob use two diagonal bases only: $\theta_\mathrm{A}=\{0,\frac{\pi}{4}\}$ for Alice, and $\theta_\mathrm{B}=\{0,\frac{\pi}{4}\}$ for Bob as well. The detections associated with identical measurement settings ($\theta_\mathrm{A}=\theta_\mathrm{B}$) are used to produce the key, while its security is guaranteed by a low enough QBER.}
\end{figure}

In both Ekert and BBM92 protocols, Alice and Bob are randomly performing measurements on a small set of bases (see Fig.~\ref{fig:protocol}). The idea it to generate a shared key by exploiting the perfect correlation of entangled states \cite{EPR35} whenever the measurements are performed in the same basis. In the BBM92 protocol \cite{BBM92} the secrecy of the key is guaranteed by a low enough quantum bit error rate (QBER). In the Ekert protocol, it is guaranteed in addition by a sufficiently large violation of Bell inequalities \cite{Bell64,CHSH,Aspect82,Weihs98,Khrennikov} measured for non-identical measurements, which allows in principle for device-independent quantum key distribution (DIQKD) \cite{Acin07,Pironio09}.

We consider an entanglement-based QKD in which the source is operated midway between Alice and Bob (see Fig.~\ref{fig:protocol}). This type of configuration allows for higher losses in optical fibers \cite{Ma07,Weihs08}, or for truly large distances in free space with a source located onboard an orbiting satellite \cite{Aspelmeyer03,Armengol08}. The source should therefore be considered to be in a location untrusted by Alice and Bob, and we will in fact assume throughout this paper that Eve is in full control of it.

\section{Multiple-photon absorption attack}
\begin{figure}
\center
\includegraphics[width=9cm]{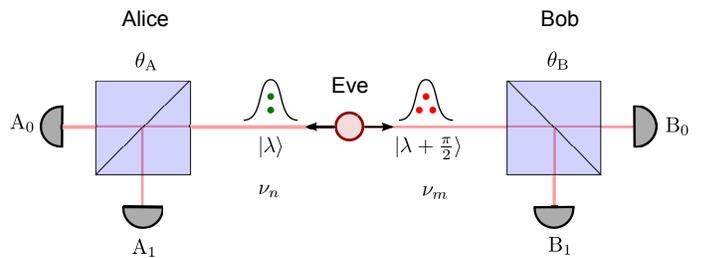}
\caption{\label{fig:attack}Principle of the Multiple-photon absorption attack. Eve sends $n$-photon pulses to Alice, the photons having a frequency $\nu_n$ and a polarization $|\lambda\rangle$; and $m$-photon pulses to Bob, the photons having a frequency $\nu_m$ and a polarization $|\lambda+\frac{\pi}{2}\rangle$. Eve chooses the frequency $\nu_n$ such that only an $n$-photon absorption is possible in Alice's detectors, and $\nu_m$ such that only an $m$-photon absorption is possible in Bob's detectors. In the figure, Eve sends 2-photon pulses to Alice, at a frequency such that a click can occur in $\mathrm{A_0}$ and $\mathrm{A_1}$ only through a two-photon absorption, and she sends 3-photon pulses to Bob, at a frequency such that a click can occur in $\mathrm{B_0}$ and $\mathrm{B_1}$ only through a three-photon absorption. }
\end{figure}

Eve's multiple-photon absorption attack \cite{Adenier10MPA} consists of replacing the source of entangled photons altogether by a controlled source of separable states (see Fig.~\ref{fig:attack}). Eve's purpose is to let Alice and Bob convince themselves that the secrecy of the key that they extract with this source is guaranteed by the laws of Quantum Mechanics, while in fact Eve has a full knowledge of the local states sent to them both, and a fairly good knowledge of the key that they generate with this source. As we will see, it does not matter which of the Ekert or BBM92 protocols is implemented by Alice and Bob. The attack works in all cases by mimicking the statistics of an entangled state well enough to pass the security checks normally undertaken by Alice and Bob: it exhibits a high violation of Bell inequalities and a low QBER.

The attack works as follows; Eve sends pulses containing photons that can each be represented by a local polarization state $|\lambda\rangle$, and she sets their frequency such that the only way to get a click in a detector is when all the photons from a pulse are absorbed in one same detector, through a multiple-photon absorption \cite{Braunstein61}. To simplify the discussion and highlight the principle of the attack, we assume that Eve can control precisely the number of photons in each pulse, their polarization and their frequency. Eve sends $n$-photon pulses to Alice, the photons having a frequency $\nu_n$ such that only an $n$-photon absorption is possible in Alice's detectors, and she sends $m$-photon pulses to Bob, the photons having a frequency $\nu_m$ such that only an $m$-photon absorption is possible in Bob's detectors. As we will see, this simple attack is sufficient to get a high enough violation of Bell inequalities and a low enough QBER as soon as $n>1$ or $m>1$.

\subsection{Single-photon absorption attack}

\begin{figure}
\center
a)
\includegraphics[width=2.95cm]{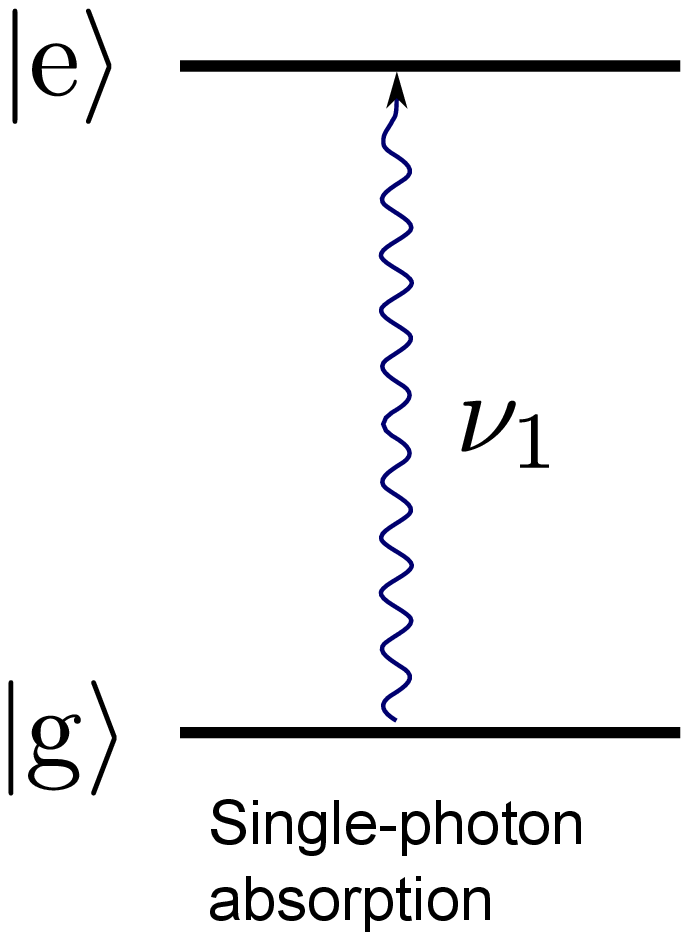}
b)
\includegraphics[width=3.5cm]{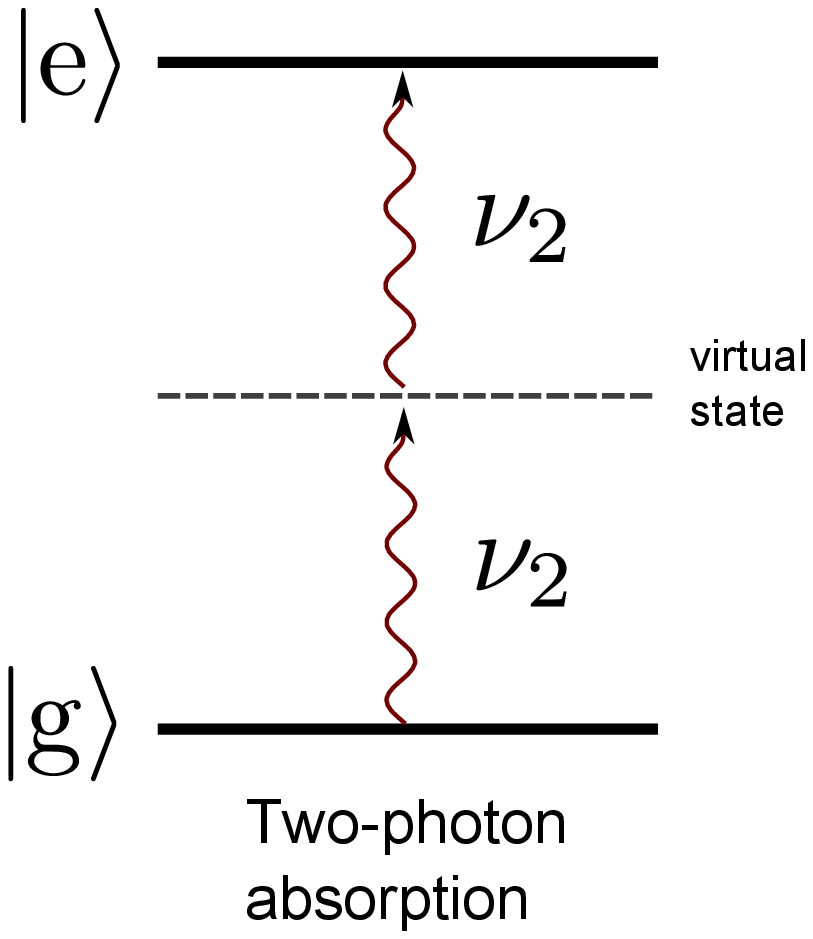}
\caption{\label{fig:absorption}Absorption of photons in the detectors: a) Single-photon absorption. b) Two-photon absorption}
\end{figure}
In this first attempt, Eve is trying to approach as best as she can the predictions of Quantum Mechanics for a singlet state, but using only a mixture of separable states, in a single-photon absorption scenario. She is therefore sending pulses such that $n=m=1$, at a frequency $\nu_1$ where the dominant absorption is a standard single-photon absorption (see Fig.~\ref{fig:absorption}-a). She will not succeed in getting a violation of Bell inequalities or a low QBER, but this result will serve as a basis to obtain a significantly better result in the case of multiple-photon absorption.

Eve prepares a set of pairs of pulses for Alice and Bob. Each pulse contains exactly one photon that is linearly polarized in a direction $\lambda_i$ chosen by Eve. It can thus be described in its own Hilbert subspace $\mathcal{H}_i$ by a two-level state of the form
\begin{equation}\label{E:10}
    |\lambda_i\rangle=\cos\lambda_i|0\rangle+\sin\lambda_i|1\rangle,
\end{equation}
where $|0\rangle$ and $|1\rangle$ are the eigenvectors of $\hat{\sigma}_\mathrm{z}$ with eigenvalues $-1$ and $+1$ respectively.
 The pair of pulses can then be described in the tensor product Hilbert space $\mathcal{H}_{12}=\mathcal{H}_1\otimes\mathcal{H}_2$ as $|\Lambda_{12}\rangle=|\lambda_1\rangle\otimes|\lambda_2\rangle$.

Alice and Bob, who are performing local measurements respectively in $\mathcal{H}_1$ and $\mathcal{H}_2$, want to measure the statistical correlation of the pairs they receive from Eve when they perform some rotations $\hat{R}(\theta_\mathrm{A})$ and $\hat{R}(\theta_\mathrm{B})$ on their respective pulses, followed by a measurement of the observable $\hat{\sigma}_\mathrm{z}$.

Since Eve wants them to obtain an as good (anti)correlation as possible with separable states, she sends orthogonal states to Alice and Bob, that is, such that $\lambda_1=\lambda$ and $\lambda_2=\lambda+\frac{\pi}{2}$, and the initial state sent to Alice and Bob can be written:
\begin{equation}\label{lambdainit}
    |\Lambda_{12}\rangle=|\lambda\rangle\otimes|\lambda+\frac{\pi}{2}\rangle
\end{equation}
For a pair initially represented by this state (\ref{lambdainit}), the state after local rotations on each sides becomes
\begin{equation}\label{lambdaAB}
    |\Lambda_\mathrm{AB}\rangle=|\lambda+\theta_\mathrm{A}\rangle\otimes|\lambda+\frac{\pi}{2}+\theta_\mathrm{B}\rangle,
\end{equation}
which can be expanded as
\begin{equation}\label{A-5}
\begin{aligned}
    |\Lambda_\mathrm{AB}\rangle=\cos(\lambda+\theta_\mathrm{A})\cos(\lambda+\frac{\pi}{2}+\theta_\mathrm{B})&|00\rangle\\
    +\sin(\lambda+\theta_\mathrm{A})\cos(\lambda+\frac{\pi}{2}+\theta_\mathrm{B})&|10\rangle\\
    +\cos(\lambda+\theta_\mathrm{A})\sin(\lambda+\frac{\pi}{2}+\theta_\mathrm{B})&|01\rangle\\
    +\sin(\lambda+\theta_\mathrm{A})\sin(\lambda+\frac{\pi}{2}+\theta_\mathrm{B})&|11\rangle,
\end{aligned}
\end{equation}

Keeping the condition of orthogonality between Alice and Bob from pair to pair, Eve is randomizing the parameter $\lambda$ associated to each pair. The state of the set of pairs prepared by Eve can therefore be described by a mixture. We can characterize the probability of obtaining a state between $|\lambda\rangle$ and $|\lambda+d\lambda\rangle$ by a probability density distribution $\rho(\lambda)$ on a single probability space $(\Omega,\mathcal{F},P)$.

The mixture $\hat{\rho}$ describing the set of pairs after rotation is therefore
\begin{equation}\label{mixture}
    \hat{\rho}_\mathrm{AB}=\int_\Omega\rho(\lambda)d\lambda \:|\Lambda_\mathrm{AB}\rangle\langle\Lambda_\mathrm{AB}|,
\end{equation}
with $\int_\Omega\rho(\lambda)d\lambda =1.$

Note that we chose to denote the polarization of the pulses prepared by Eve as $\lambda$, a symbol that is traditionally reserved to hidden-variables, but this variable is in fact hidden only from Alice and Bob, and it is not a hypothetical supplementary parameters to Quantum Theory. By describing the state with mixtures, we adopt here the point of view of Alice and Bob. However, the parameter $\lambda$ describing the state of each pulse is freely chosen by Eve. Thus it can in principle even be deterministic, as a function of time. Eve needs only to make sure that there is on average no preferred direction for the $\lambda$ so that the source is rotationally invariant for Alice and Bob, both at the single-count and coincidence-count level.

After performing their respective rotation on the received pulses, Alice and Bob perform a joint measurement $\sigma_\mathrm{z}\otimes\sigma_\mathrm{z}$ on each pair. The expectation value of this measurement is:
\begin{equation}\label{A-7}
    E_\mathrm{AB}=\langle\hat{\sigma}_\mathrm{z}\otimes\hat{\sigma}_\mathrm{z}\rangle_{\hat{\rho}_\mathrm{AB}}=\textrm{Tr}(\hat{\rho}_\mathrm{AB}\hat{\sigma}_\mathrm{z}\otimes\hat{\sigma}_\mathrm{z}),
\end{equation}
with $\hat{\sigma}_\mathrm{z}=|0\rangle\langle0|-|1\rangle\langle1|$, so that
$$\hat{\sigma}_\mathrm{z}\otimes\hat{\sigma}_\mathrm{z}=|00\rangle\langle00|-|10\rangle\langle10|-|01\rangle\langle01|+|11\rangle\langle11|,$$
and where $\textrm{Tr}$ is the trace, defined as the sum of the diagonal matrix elements in any orthonormal basis, which in our case is $\textrm{Tr}\:\hat{O}=\sum_{i,j=0}^1\langle ij|\hat{O}|ij\rangle$, where $\hat{O}$ is an operator in $\mathcal{H}_{12}$.

By linearity of the trace, we can write
$$E_\mathrm{AB}=P_{00}-P_{10}-P_{01}+P_{11},$$
with
$$P_{kl}=\textrm{Tr}(\hat{\rho}_\mathrm{AB}|kl\rangle\langle kl|),$$
where $k$ and $l$ are 0 or 1.

The $P_{kl}$ are the joint probabilities, and we can now express them explicitly, using Eq.~(\ref{mixture}), as
$$P_{kl}=\sum_{i,j=0}^1\int_\Omega\rho(\lambda)d\lambda \: \langle ij|\Lambda_\mathrm{AB}\rangle\langle\Lambda_\mathrm{AB}|kl\rangle\langle kl|ij\rangle,$$
which, since $\{|00\rangle,|10\rangle,|01\rangle,|11\rangle\}$ forms an orthonormal basis in $\mathcal{H}_{12}$, simplifies as
\begin{equation}\label{Pkl}
    P_{kl}=\int_\Omega\rho(\lambda)d\lambda \:\big|\langle kl|\Lambda_\mathrm{AB}\rangle\big|^2.
\end{equation}

Note that, owing to the separability of $|\Lambda_\mathrm{AB}\rangle$ as expressed by Eq. (\ref{lambdaAB}), we can factorize the integrand in the above joint probability into a product, that is,
\begin{equation}\label{factorPkl}
    P_{kl}=\int_\Omega\rho(\lambda)d\lambda \:\big|\langle k|\lambda+\theta_\mathrm{A}\rangle \langle l|\lambda+\frac{\pi}{2}+\theta_\mathrm{B}\rangle\big|^2.
\end{equation}

Denoting the probability to get a photon in a channel $i$ for a local state $|\lambda\rangle$ and a measurement direction $\theta$ as
\begin{equation}\label{psinglelocal}
    P_i(\lambda,\theta)=\big|\langle i|\lambda+\theta\rangle\big|^2,
\end{equation}
which is explicitly
\begin{equation}\label{Malus}
    \begin{aligned}
    P_0(\lambda,\theta)=\cos^2(\lambda+\theta),\\
    P_1(\lambda,\theta)=\sin^2(\lambda+\theta),
    \end{aligned}
\end{equation}
we can rewrite Eq.~(\ref{factorPkl}) as
\begin{equation}\label{jointprobexp}
P_{kl}=\int_\Omega\rho(\lambda)d\lambda \; P_k(\lambda,\theta_\mathrm{A})\; P_l(\lambda+\frac{\pi}{2},\theta_\mathrm{B}),
\end{equation}
which can be simply interpreted as the integral on all possible $\lambda$ of the product of the probability to detect a photon in channel $k$ on Alice's side for a local state $|\lambda\rangle$ by the probability to get a photon in channel $l$ on Bob's side for a local state $|\lambda+\frac{\pi}{2}\rangle$.

Assuming $\rho(\lambda)$ is a uniform distribution on the interval $[0,2\pi[$, we can then express these integrals explicitly as
\begin{equation}\label{jointprob}
    \begin{aligned}
P_{00}&=\frac{1}{2\pi}\int_0^{2\pi}d\lambda \cos^2(\lambda+\theta_\mathrm{A})\cos^2(\lambda+\frac{\pi}{2}+\theta_\mathrm{B}),\\
P_{10}&=\frac{1}{2\pi}\int_0^{2\pi}d\lambda \sin^2(\lambda+\theta_\mathrm{A})\cos^2(\lambda+\frac{\pi}{2}+\theta_\mathrm{B}),\\
P_{01}&=\frac{1}{2\pi}\int_0^{2\pi}d\lambda \cos^2(\lambda+\theta_\mathrm{A})\sin^2(\lambda+\frac{\pi}{2}+\theta_\mathrm{B}),\\
P_{11}&=\frac{1}{2\pi}\int_0^{2\pi}d\lambda \sin^2(\lambda+\theta_\mathrm{A})\sin^2(\lambda+\frac{\pi}{2}+\theta_\mathrm{B}),
    \end{aligned}
\end{equation}
which leads to
\begin{equation}\label{jointp}
    \begin{aligned}
    P_{00}&=P_{11}=\frac{1}{8} (2 - \cos 2(\theta_\mathrm{A} -\theta_\mathrm{B})),\\
    P_{10}&=P_{01}=\frac{1}{8} (2 + \cos 2(\theta_\mathrm{A} -\theta_\mathrm{B})),
    \end{aligned}
\end{equation}
and finally, we get
\begin{equation}\label{corrth}
    E_\mathrm{AB}=-\frac{1}{2}\cos2(\theta_\mathrm{A}-\theta_\mathrm{B}).
\end{equation}

This result differs essentially from the prediction for the singlet state by its visibility of 1/2 instead of 1. With this correlation function, the maximum of the CHSH function \cite{CHSH}, defined as
\begin{equation}\label{CHSH}
    S=|E(\theta_\mathrm{A},\theta_\mathrm{B})
    +E(\theta^\prime_\mathrm{A},\theta_\mathrm{B})
    +E(\theta_\mathrm{A},\theta^\prime_\mathrm{B})
    -E(\theta^\prime_\mathrm{A},\theta^\prime_\mathrm{B})|
\end{equation}
is $S=\sqrt{2}$, which is clearly below 2 and therefore insufficient as an attack on Ekert protocol.

Similarly, this attack fails against the BBM92 protocol. The QBER can be estimated \cite{Scheidl09,Branciard05} from the visibility $\mathrm{V}$ as
$\mathrm{QBER}=\frac{1-\mathrm{V}}{2}$. With $\mathrm{V}=1/2$, the QBER measured by Alice and Bob is as high as $25\%$ in this single-photon attack, which Alice and Bob would not fail to reject.

\subsection{Two-photon absorption attack}
In the two-photon absorption attack, Eve similarly sends pairs of pulses described by a mixture of separable states (\ref{mixture}) to Alice and Bob, but this time each pulse contains two photons instead of one. The photons inside a pulse share the same state: $|\lambda\rangle$ for the two photons inside the pulse sent to Alice; and $|\lambda+\frac{\pi}{2}\rangle$ for the two photons inside the pulse sent to Bob. We assume that the photons inside a pulse are independent: each photon follows the rules of Quantum Mechanics as prescribed by its quantum state independently of what the other photons inside the pulse are doing.

Now, the crucial difference is that these photons are chosen by Eve with a frequency $\nu_2$ lower than in the single-photon absorption case discussed above, so that the energy of a single photon is insufficient to trigger a click. Eve chooses the frequency of the photons such that the only way to get a click is through a two-photon absorption (see Fig.~\ref{fig:absorption}-b). We assume for simplicity that whenever two photons hit the same detector simultaneously, the probability that they trigger a click in the detector is 1. As we will see, this feature alone is (surprisingly) enough to lead to a clear violation of Bell inequalities and a low QBER.

So, a click occurs in a specific output channel of a polarizing beam-splitter (PBS) only when the two photons inside the same pulse choose to exit through that same channel. If they choose different channels, no two-photon absorption can occur because there is only one photon in each channel. On each side, the three possibilities are:
\begin{itemize}
  \item Both photons go to channel 0 $\rightarrow$ click in channel 0 through a two-photon absorption,
  \item Both photons go to channel 1 $\rightarrow$ click in channel 1 through a two-photon absorption,
  \item One photon goes to channel 0, the other goes to channel 1 $\rightarrow$ no click in either channel.
\end{itemize}
Note that this third possibility brings us in the realm of the detection loophole \cite{Pearle,Larsson98,Gisin99,Adenier08,Adenier09}. It is the essential reason for the appearance of the violation of Bell inequalities that Alice and Bob are going to obtain. It should however be stressed that the non detections come from the frequency threshold in the photoelectric effect alone; a feature that is relevant in all detectors based on this effect, regardless of their quantum efficiency. Eve is therefore working in a fully Quantum Mechanical framework, without assuming anything about the detectors other than the existence of two-photon absorption processes at certain frequencies, and without assuming the existence of any hidden-variables.

We want to calculate the probabilities $P_{kl}^{(2)}$ that Alice and Bob get a coincidence click respectively in channels $k$ and $l$ in a two-photon absorption attack.

Note that whenever it is necessary to avoid possible ambiguities, we will label hereafter the equations with an upper script $(n)$ or $(n,m)$ indicating the order of the multiple-photon attack. For instance the joint probability $P_{kl}^{(2)}$ is for a two-photon absorption attack, while $P_{kl}^{(2,3)}$ is the same probability for a mixed attack in which Alice is subjected to a two-photon attack while Bob is subjected to a three-photon attack.

Assuming independence between the photons, for a local state $|\lambda\rangle$ and a measurement angle $\theta$ on either side, the probability that both photons from one pulse end up in the same channel $i$ is simply the square of the probability $P_i(\lambda,\theta)$ to see one such photon exit the PBS through this channel $i$, that is: $P_i^2(\lambda,\theta)$

So, similarly to what we had in the single photon case in Eq.(\ref{jointprobexp}), the probability $P_{kl}^{(2)}$ to get a click in channel $k$ for Alice and in channel $l$ for Bob in a two-photon absorption process is therefore the integral over all possible state of the product of the probability $P_k^2(\lambda,\theta_\mathrm{A})$ for Alice to get a click in channel $k$ by the probability $ P_l^2(\lambda+\frac{\pi}{2},\theta_\mathrm{B})$ for Bob to get a click in channel $l$:
\begin{equation}\label{jointprobexptwo}
P_{kl}^{(2)}=\int_\Omega\rho(\lambda)d\lambda \; P_k^2(\lambda,\theta_\mathrm{A})\; P_l^2(\lambda+\frac{\pi}{2},\theta_\mathrm{B}).
\end{equation}

For a rotationally invariant source, $\lambda$ is uniformly distributed on the interval $[0,2\pi[$, which leads to:
\begin{equation}\label{jointprobtp}
    \begin{aligned}
P_{00}^{(2)}&=\frac{1}{2\pi}\int_0^{2\pi}d\lambda \: \cos^4(\lambda+\theta_\mathrm{A})\cos^4(\lambda+\frac{\pi}{2}+\theta_\mathrm{B}),\\
P_{10}^{(2)}&=\frac{1}{2\pi}\int_0^{2\pi}d\lambda \: \sin^4(\lambda+\theta_\mathrm{A})\cos^4(\lambda+\frac{\pi}{2}+\theta_\mathrm{B}),\\
P_{01}^{(2)}&=\frac{1}{2\pi}\int_0^{2\pi}d\lambda \: \cos^4(\lambda+\theta_\mathrm{A})\sin^4(\lambda+\frac{\pi}{2}+\theta_\mathrm{B}),\\
P_{11}^{(2)}&=\frac{1}{2\pi}\int_0^{2\pi}d\lambda \: \sin^4(\lambda+\theta_\mathrm{A})\sin^4(\lambda+\frac{\pi}{2}+\theta_\mathrm{B}),
    \end{aligned}
\end{equation}
and we obtain:
\begin{equation}
    \begin{aligned}
P_{00}^{(2)}&=P_{11}^{(2)}=\frac{1}{128} (18 - 16 \cos 2(\theta_\mathrm{A} - \theta_\mathrm{B}) + \cos 4 (\theta_\mathrm{A} - \theta_\mathrm{B}),\\
P_{01}^{(2)}&=P_{10}^{(2)}=\frac{1}{128} (18 + 16 \cos 2(\theta_\mathrm{A} - \theta_\mathrm{B}) + \cos 4 (\theta_\mathrm{A} - \theta_\mathrm{B}).
    \end{aligned}
\end{equation}

Note that these four probabilities no longer add up to 1, because of the cases involving a non detection on either side or both, which are discarded by Alice and Bob. So, as is standard in optical EPR experiment and in entanglement-based QKD with photons, the correlation function has to be normalized by the sum $\sum_{k,l}P_{kl}^{(2)}$:
\begin{equation}
E^{(2)}_\mathrm{AB}=
\frac{P_{00}^{(2)}-P_{10}^{(2)}-P_{01}^{(2)}+P_{11}^{(2)}}
{P_{00}^{(2)}+P_{10}^{(2)}+P_{01}^{(2)}+P_{11}^{(2)}}.
\end{equation}

We obtain explicitly
$$E^{(2)}_\mathrm{AB}=-\frac{16 \cos 2(\theta_\mathrm{A} - \theta_\mathrm{B})}{18+\cos 4(\theta_\mathrm{A} - \theta_\mathrm{B})},$$
which lead to a violation of Bell inequalities for $\theta_\mathrm{A}=\{0,\frac{\pi}{4}\}$ and $\theta_\mathrm{B}=\{-\frac{\pi}{8},\frac{\pi}{8}\}$ of
$$S^{(2)}=\frac{16}{18}\;2\sqrt{2}\approx 2.51,$$
which is clearly above 2.

Here the visibility is $V=0.842$, so that a BBM92 protocol would give a QBER of $7.9\%$, which is already well below the security bound of $11\%$ against coherent attacks \cite{Branciard05,Ma07,Weihs08}.

\subsection{Multiple-photon absorption attack}
A similar demonstration in the case of a three-photon absorption, with Eve sending three photons per pulse, leads to
\begin{equation}\label{jointprobthree}
P_{kl}^{(3)}=\int_\Omega\rho(\lambda)d\lambda \; P_k^3(\lambda,\theta_\mathrm{A})\; P_l^3(\lambda+\frac{\pi}{2},\theta_\mathrm{B})
\end{equation}
which exhibits a violation of Bell inequalities as high as $S^{(3)}\approx 3.17$, and a QBER of $2.1\%$. Quite generally, a multi-photon absorption process can lead to a violation of Bell inequalities as large as desired within the algebraic limit of 4, the only limit being the order of the multiple-photon absorptions that Eve can drive into Alice's and Bob's detectors.

Note that if Eve sends pulses meant to drive on one side a two-photon absorption and on the other a three-photon absorption (a feature that Eve could achieve by alternatively sending photons of different frequency to Alice and Bob), the relevant probabilities for the coincidences are of the form
\begin{equation}\label{jointprobexpthree}
P_{kl}^{(2,3)}=\int_\Omega\rho(\lambda)d\lambda \; P_k^2(\lambda,\theta_\mathrm{A})\; P_l^3(\lambda+\frac{\pi}{2},\theta_\mathrm{B})
\end{equation}
which leads to a correlation
$$E^{(2,3)}_\mathrm{AB}=-\frac{10 \cos 2(\theta_\mathrm{A} - \theta_\mathrm{B})}{10+\cos 4(\theta_\mathrm{A} - \theta_\mathrm{B})}$$
and a violation of Bell inequalities for $\theta_\mathrm{A}=\{0,\frac{\pi}{4}\}$ and $\theta_\mathrm{B}=\{-\frac{\pi}{8},\frac{\pi}{8}\}$ of exactly $$S^{(2,3)}=2\sqrt{2},$$
which is the maximum violation predicted by Quantum Mechanics for entangled states.

In case of a BBM92 protocol, the QBER measured by Alice and Bob is $4.5\%$. This is below the critical QBER of $7.1\%$ for a device-independent quantum key distribution (DIQKD) in which not only the source but also the measurement devices used by Alice and Bob are untrusted \cite{Acin07,Pironio09}. The reason for this behavior is essentially that the detection probability of the pulses is below the $83\%$ required to close the detection loophole \cite{GargMermin,Larsson98}. This shows the absolute necessity of a loophole-free violation of Bell inequalities in a DIQKD scenario \cite{Acin07,Pironio09}.

\begin{figure}
\center
\includegraphics[width=8cm]{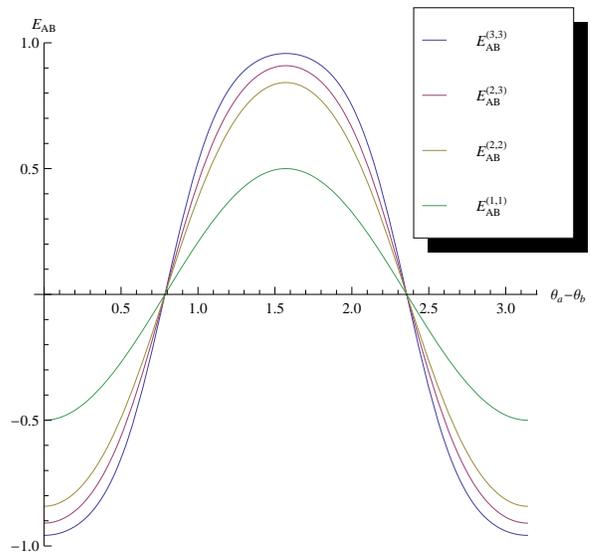}
\caption{\label{fig:correlations}Correlations measured by Alice and Bob depending on the order of Eve's multiple-photon absorption attack. An important property of this attack is that the correlations measured by Alice and Bob are rotationally invariant: they only depend on the difference between their measurement angles $\theta_\mathrm{A}$ and $\theta_\mathrm{B}$, and not on their absolute values. It means that the attack is basis independent, and that it works independently of which entanglement-based QKD protocol is chosen by Alice and Bob.}
\end{figure}

The performances of the various multiple-photon absorption attacks are summarized in Table~\ref{table1}, and the corresponding correlation functions measured by Alice and Bob are displayed in Fig.~\ref{fig:correlations}.

\subsection{Multiple-photon absorption attack as an adaptive process}
In order to interpret and explain this seemingly unlikely violation of Bell inequalities with a mixture of separable states, it is convenient to look locally (e.g. on Alice's side) at the probability that a pulse $\lambda$ impinging on a polarizing beam-splitter gets detected in either output channel.

In the multiple-photon absorption attack, the only way to get a click with a pulse containing $n$ photon is through an $n$-photon absorption, so that the probability to get a click in an output channel $i$ is equal to the probability that all the photons from a pulse exit through this same channel. Assuming the independence of the photons, this probability is simply the power $n$ of the probability $P_i(\lambda,\theta)$ of Eq.~(\ref{psinglelocal}) that a single photon polarized along $\lambda$ goes to channel $i$. The probability to get a click in either output channel for an incoming $n$-photon pulse is therefore:
\begin{equation}\label{pclic}
\begin{aligned}
    P^{(n)}_\mathrm{clic}&=\big(P_0(\lambda,\theta)\big)^n+\big(P_1(\lambda,\theta)\big)^n\\
    &=\cos^{2n}(\lambda+\theta)+\sin^{2n}(\lambda+\theta).
\end{aligned}
\end{equation}

In the single-photon attack presented in the first section ($n=1$), each pulse polarized along $\lambda$ contains one photon only, and the probabilities to get a click in either channel are adding up to one, that is, $P^{(1)}_\mathrm{clic}=1$, which denotes that a photon impinging on a PBS is necessarily detected in one of the output channels in the ideal case. All the single-photon pulses are therefore treated on equal footing, independently of the state $\lambda$ and of the measurement setting $\theta$, and this is why this attack fails to lead to an observed violation of Bell inequalities.

\begin{figure}
a)
  \includegraphics[width=9cm]{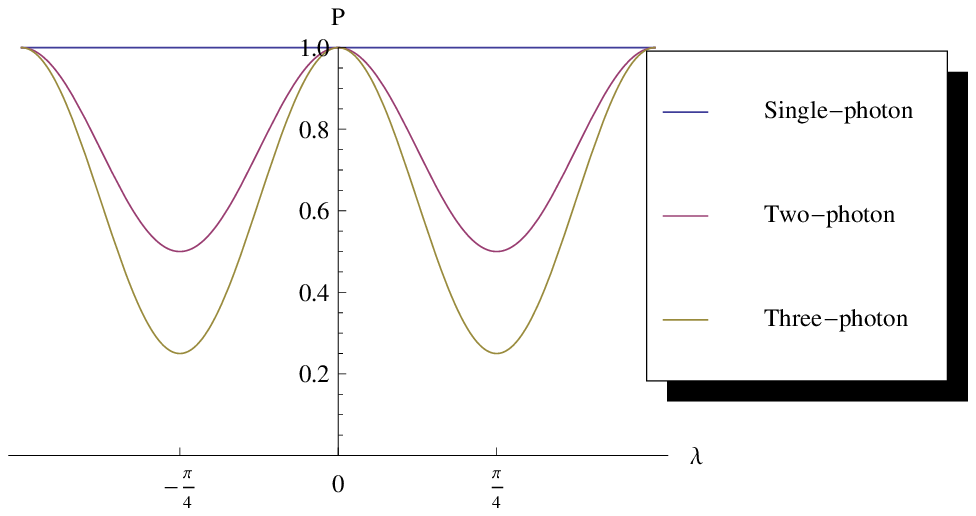}\\
b)
  \includegraphics[width=9cm]{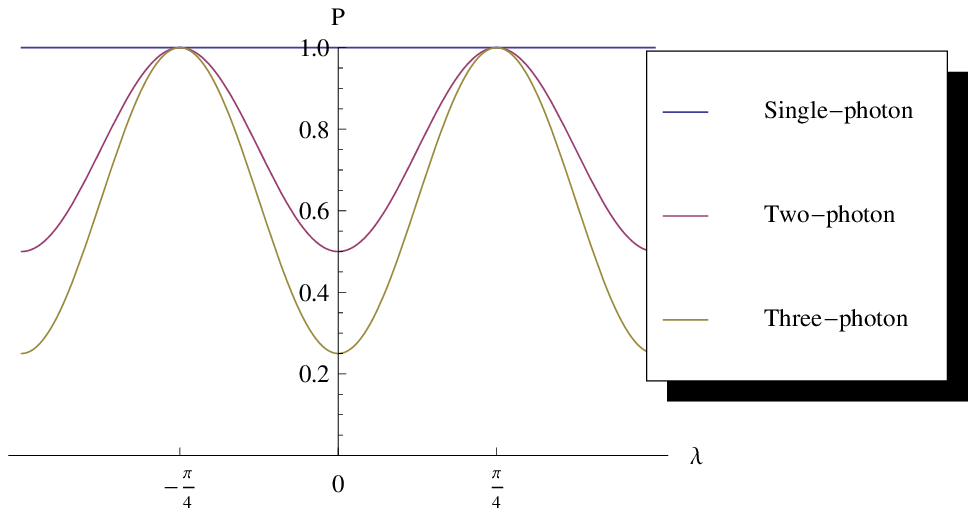}\\
  \caption{Probability of detection of a pulse $\lambda$ in either output channel: a) for $\theta=0$, b) for $\theta=\pi/4$. In the single-photon case, the probability to detect a pulse is identically equal to $1/2$, which denotes a passive response of the measurement apparatus: all single-photon pulses are treated on equal ground. By contrast, in the two-photon and three-photon absorption attacks the process is adaptive in the sense that the probability that a pulse $\lambda$ gets detected in either channel depends strongly on the measurement context $\theta$ encountered.}\label{figpclic}
\end{figure}

By contrast, in the case of a two-photon absorption attack, the probability that a pulse containing $n$ photons polarized along $\lambda$ gets detected in either channel becomes
\begin{equation}\label{pclic2}
\begin{aligned}
    P^{(2)}_\mathrm{clic}&=\cos^{4}(\lambda+\theta)+\sin^{4}(\lambda+\theta)\\
    &=\frac{1}{4}(3+\cos(4(\lambda+\theta)).
\end{aligned}
\end{equation}
It exhibits a clear dependence on the state of the pulse $\lambda$ and the local measurement setting $\theta$. It means that the interaction of each pulse $\lambda$ with the measurement apparatus as a whole (constituted by the polarizing beam-splitter and the detectors) depends on the very setting $\theta$ of this apparatus: the detection process in the multiple-photon absorption attack is \emph{state-adaptive} \cite{Ohya}. This behavior is illustrated in Fig.~\ref{figpclic}. In the two-photon and three-photon absorption attacks, the response of the measurement apparatus to incoming pulses is radically different when the measurement setting is $\theta=0$ than when it is $\theta=\frac{\pi}{4}$. This behavior is hidden from Alice and Bob by the randomness of $\lambda$ from one pair to the other, but the consequence is that different parts of the probability space are in effect weighted as a function of the local context $\theta$. It leads to a bi-local dependence on the contexts $(\theta_\mathrm{A},\theta_\mathrm{B})$ when calculating the probabilities associated with the coincidences, and this feature is known to open the possibility of a violation of Bell inequalities with local states \cite{Ohya,Accardi,Khrennikov}.

\subsection{Eve's knowledge of the key}
In a multiple-photon absorption attack, Eve has a full knowledge of the local states that she sends to Alice and Bob. Each pair is characterized by state of the form Eq.(\ref{lambdainit}), where the polarization $\lambda$ characterizing each pair is chosen by Eve. This perfect knowledge of the local state sent to Alice and Bob does not however mean that Eve has a perfect knowledge of the key that they will generate with these pairs, because of the inherent probabilistic nature of Quantum predictions whenever one performs a measurement in another basis than the one in which a state was prepared.

The key is generated in those cases in which Alice and Bob are performing identical measurement, that is $\theta_\mathrm{A}=\theta_\mathrm{B}=\theta$. Knowing $\alpha=|\lambda+\theta|$ for each pulse, Eve knows the probabilities for 0 and 1 to be realized on each side, and she simply bets for the one that has the higher probability.

Consider the cases for which $0<|\lambda+\theta|<\pi/4$. We then have $\cos^{2}(\lambda+\theta)>\sin^{2}(\lambda+\theta)$, so that $P_0(\lambda,\theta)>P_1(\lambda,\theta)$ for Alice and $P_1(\lambda+\pi/2,\theta)>P_0(\lambda+\pi/2,\theta)$ for Bob. Eve therefore bets that the bit measured by Alice is 0 and that the bit measured by Bob is 1.

Assuming independence between the photons, the probability that all the photons from an $n-$photon pulse on Alice's side end up in channel $1$, and thus generate a click opposite to Eve's guess, is  $P_1^n(\lambda,\theta)$. Just the same, the probability that all the photons from an $m-$photon pulse on Bob's side end up in channel $0$, and thus generate a click opposite to Eve's guess, is  $P_0^m(\lambda+\pi/2,\theta)$. The probability that Eve guesses incorrectly a bit shared by Alice and Bob when $0<|\lambda+\theta|<\pi/4$ is therefore
$$P_1^n(\lambda,\theta)P_0^m(\lambda+\pi/2,\theta)=\sin^{2(n+m)}(\lambda+\theta).$$

Similarly, in the cases for which $\pi/4<|\lambda+\theta|<\pi/2$, the probability that Eve guesses incorrectly is
$$P_0^n(\lambda,\theta)P_1^m(\lambda+\pi/2,\theta)=\cos^{2(n+m)}(\lambda+\theta).$$

The same reasoning can be extended to higher values of $|\lambda+\theta|$, leading either to a dependence on $\sin^{2(n+m)}(\lambda+\theta)$ or $\cos^{2(n+m)}(\lambda+\theta)$, so that on average, for a uniform distribution of $\lambda$ on the circle, the probability that Eve guesses incorrectly a bit shared by Alice and Bob is
$$P_\mathrm{error}^{(n,m)}=\frac{2}{\pi}\big(\int_0^{\pi/4}d\alpha \: \sin^{2(n+m)}\alpha+\int_{\pi/4}^{\pi/2}d\alpha \: \cos^{2(n+m)}\alpha\big).$$

Note that the pulses for which Eve has the least information are those where $\cos^{2n}(\lambda+\theta)\approx\sin^{2n}(\lambda+\theta)$, that is $|\lambda+\theta|\approx \pi/4+k\pi/2$. It is also for these pulses that Alice and Bob are the most likely to not end up with the same bit, thus increasing the measured QBER. Unfortunately for Alice and Bob, as soon as $n>1$ or $m>1$, it is also for these pulses that the probability of detection is the smallest for a fixed $n$ and $m$, as given by Eq.~(\ref{pclic}). Since this probability decreases for higher-order multiple-photon absorption, it also means that, rather counter-intuitively, the probability $P_\mathrm{error}^{(n,m)}$ that Eve makes a mistake guessing a bit shared by Alice and Bob decreases with higher violation of Bell inequality and with lower QBER (see Table \ref{table1}).

Alice and Bob should therefore be wary not to trust a strong violation of Bell inequality and/or a low QBER as such, because it might give them the wrong impression that the strength of this violation and/or the small number of errors in the key make it safe when it is in fact the opposite.

\begin{table}
  \caption{Performances of multiple-photon absorption attacks. $\eta$ is the minimum channel efficiency required to rule out each attack; $S$ is the violation of the CHSH inequality measured by Alice and Bob; $\mathrm{V}$ is the visibility; QBER is the quantum bit error rate; and $P_\mathrm{error}^{(n,m)}$ is the corresponding probability for Eve to incorrectly guess a bit shared by Alice and Bob.}
\begin{center}
\begin{tabular}{|c|c|c|c|c|c|}
  \hline
  (n,m) & $\eta$ & S & V & QBER & $P_\mathrm{error}^{(n,m)}$ \\
  \hline
  (1,1) & $100\%$ & $\sqrt{2}$ & 0.50 & $25\%$ & $5.67\%$ \\
  \hline
  (2,2) & $75\%$ & 2.51 & $0.84$ & $7.9\%$ & $0.82\%$ \\
  \hline
  (2,3) & $68.75\%$ & $2\sqrt{2}$ & $0.91$ & $4.5\%$ & $0.17\%$ \\
  \hline
  (3,3) & $62.5\%$ & 3.17 & $0.96$ & $2.1\%$ & $0.14\%$\\
  \hline
\end{tabular}\label{table1}
\end{center}
\end{table}

\section{Countermeasures}
\subsection{Monitoring the single and coincidence counts}
Detecting Eve's attack by monitoring closely the single counts and the coincidence counts is not a trivial task.

As we have seen, a violation of Bell inequalities or a low QBER is not a trustworthy criterion unless the detection efficiency is higher than the values given in Table \ref{table1}. If Alice and Bob were taking into account the non-detected pulses to compute the correlation, for instance by assigning a random bit value to non-detected pulses \cite{Acin07}, they would not get any violation of Bell inequality with this attack. However, the typical detection efficiencies obtained in actual implementation of QKD protocols with photons are far below the bounds given in Table \ref{table1}. Eve's attack is therefore relevant because discarding the pairs for which no detection is recorded on either side and normalizing by the sum of coincidences is the standard way of dealing with non-detections. Alice and Bob would therefore not be able to distinguish a genuine source of entangled photons from this attack by simply observing a violation of Bell inequalities.

Monitoring the single counts is also unlikely to betray Eve's attack, because all the channels are treated on equal footing by the attack, and the source is rotationally invariant. The channel efficiencies are therefore balanced: the marginals are random for each measurement, as would be expected from a singlet state.

In the ideal case, the multiple-photon absorption attack does not produce any double-counts either that could be spotted by Alice and Bob, because a click occurs only when all the photons from a pulse exit through the same channel. In case of a two-photon absorption attack for instance, the only way to get a double-count would be with a pulse containing no less than 4 photons, which is impossible once we assume that Eve controls the number of photons in each pulse. Even if, more realistically, Eve had not a tight control on the number $n$ of photons inside a pulse, for instance if she was using a weak coherent source for which the distribution of the number of photons is Poissonian $P(n)=e^{-\mu}\frac{\mu^n}{n!}$, a double-count would only be possible for pulses containing 4 photons or more, which would have a much smaller probability to occur than the probability to get a pulse with 2 photons, for small enough values of $\mu$.

Looking at the individual rates of coincidences is also unlikely to betray Eve's attack. The predicted rates of coincidences and the correlations functions are all rotationally invariant: they depend only on the angle difference $|\theta_\mathrm{B}-\theta_\mathrm{B}|$, as is the case for a genuine singlet state. The attack is basis independent.

The correlation function differs slightly from the $-\mathrm{V}\cos 2(\theta_\mathrm{A} - \theta_\mathrm{B})$ expected for a singlet state in a lossy channel, but this difference is small and in actuality it would be difficult to distinguish the correlations obtained with a multiple-photon absorption attack from the correlation predicted for a genuine entangled state with a reduced visibility, especially if only a few points of the correlation are actually measured, as is the case in an entanglement-based QKD protocol.

An unwanted feature that could betray Eve's presence is that the sum of coincidences depends on the measurement settings $\theta_\mathrm{A}$ and $\theta_\mathrm{B}$. The stronger the violation of Bell inequalities, the stronger the visibility of the sum of coincidences. For instance, its visibility is about 0.06 in the two-photon absorption case ($S^{(2)}\approx 2.51$), and it is 0.10 in the mixed case with two-photon absorption on one side and three-photon on the other side ($S^{(2,3)}=2\sqrt{2}.$). However, Eve can remove this unwanted effect entirely by driving different detection patterns for Alice and Bob, in a similar way to what was done by Larsson \cite{Larsson98} and Gisin \cite{Gisin99} in their hidden-variable models. The simplest method would be to alternatively drive a single photon absorption on one side, and a multiple-photon absorption on the other side. The sampling is then always fair on the side driven to a single photon absorption, and the total number of coincidences becomes independent of the measurement angles. It is nevertheless in Alice's and Bob's interest to closely monitor the sum of coincidence for any such angle dependence, because if it does not guarantee in principle against this attack, it makes Eve's task more difficult by forcing her to use higher order multiple-photon absorptions to achieve the same violation of Bell inequalities or the same QBER.

\subsection{Precluding multiple-photon absorption attacks}
The assumption of fair sampling \cite{Adenier07,Adenier08,Adenier09} that is considered to be reasonable for experiments constraining the possible models of Nature \cite{Aspect82,Weihs98} is made invalid by the multiple-photon absorption attack: the sample of detected pulses does not represent fairly the pulses that are emitted. The fair sampling assumption becomes equivalent in this case to the assumption that under the conditions of use in an actual entanglement-based QKD, the detectors are actually sensitive to the single-photons received by Alice and Bob.

As reasonable as this assumption might be in the ideal case, Alice and Bob would still need to be very careful to ensure that the imperfections in the various measuring devices are not compromising its validity in actual implementations of QKD protocols.

Using frequency filters so that only a known range of restricted frequencies can reach the detectors is an obvious way to limit the possibility of multiple-photon absorption in the ideal case. However, in actual implementations of QKD protocols, Alice and Bob still need to guarantee that the dominant way to get a click in their necessarily imperfect detector is through a single-photon absorption, even at those frequencies allowed by the filters.

The energy diagram displayed on figure \ref{fig:absorption} is that of an ideal two-level system, so that the frequency at which a two-photon absorption occurs is exactly half that of a single-photon absorption. In a real detector however, where the energy levels are bound to be more complicated, photons with a frequency just slightly lower than $\nu_1$ could trigger a two-photon absorption if there exist other stable energy levels above $|\mathrm{g}\rangle$.

This issue becomes far more critical considering the existing faked-state attacks that have already been successfully implemented against QKD protocols \cite{Makarov07,Makarov10a,Makarov10}, by forcing the detectors to exit the single-photon sensitive Geiger mode \cite{Makarov10a}. It is conceivable that this type of blinding attack could be tailored to allow only a multiple-photon absorption in the detectors used by Alice and Bob, at those precise frequencies that are permitted to go through. Such a combination of blinding detector attack together with our source designed for a multiple-photon absorption attack could then constitute a robust attack even against Ekert protocol, something that the blinding attack alone could not do \cite{Makarov07}.

\subsection{Testing the fairness of the sample}
An active method for Alice and Bob to detect Eve's attack, or to make sure that reducing the dark counts is not made at the cost of increasingly biasing the detection of signal pulses, would be to implement a fair sampling test \cite{Adenier08b} adapted to quantum key distribution with entangled states \cite{Adenier10FS}.

This fair sampling test does not introduce any loss and can be performed locally and unilaterally on either side during the production of the key. In a nutshell, it consists of analyzing the output channels of the measuring devices (here, the polarizing beam-splitters) instead of simply feeding detectors with them. The standard design of an entanglement-based QKD is kept intact, with two polarizing beam-splitters on each side (Alice and Bob) projecting the incoming pulses on random bases $\theta_\mathrm{A}$ and $\theta_\mathrm{B}$, as depicted on Fig.~\ref{fig:protocol}. The novelty is to replace each detector by a polarimeter: a polarizing beam-splitter followed by a detector at each output.

\begin{figure}
\center
\includegraphics[width=8cm]{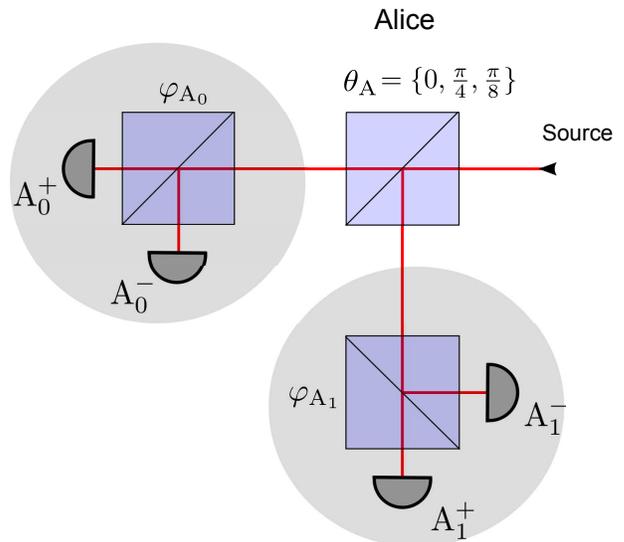}
\caption{\label{fig:fstest10}Fair Sampling test on Alice's side. The detector in channel $0$ is replaced by a polarimeter $A_0$ with two detectors $\mathrm{A}_0^+$ and $\mathrm{A}_0^-$. The detector in channel $1$ is replaced by a polarimeter $\mathrm{A}_1$ with two detectors $\mathrm{A}_1^+$ and $\mathrm{A}_1^-$. Ekert protocol is thus unaltered by the fair sampling test: polarimeter $\mathrm{A}_0$ is equivalent to the detector in channel $0$ in Fig.~\ref{fig:protocol}, and polarimeter $\mathrm{A}_1$ is equivalent to the detector in channel $1$. Bob would proceed similarly to test the fairness of the sampling on his side.}
\end{figure}

Consider Alice's side (see Fig. \ref{fig:fstest10}). We label the polarimeter in channel $0$ as $\mathrm{A}_0$, the orientation of its polarizing beam-splitter as $\varphi_\mathrm{A_0}$, and the detectors in the transmitted and reflected output as $\mathrm{A}_0^+$ and $\mathrm{A}_0^-$ respectively. A click in either of these two detectors is treated as a click in channel 0.

Similarly, the polarimeter in channel $1$ is labeled $\mathrm{A}_1$, the orientation of its polarizing beam-splitter $\varphi_\mathrm{A_1}$, and the detectors in the transmitted and reflected output are $\mathrm{A}_1^+$ and $\mathrm{A}_1^-$ respectively. A click in either of these two detectors is treated as a click in channel 1.

Bob would proceed similarly with two polarimeters labeled $\mathrm{B}_0$ and $\mathrm{B}_1$.

Let us denote $P_\mathrm{0,+}$ the probability that a photon initially polarized along $\lambda$ ends up in the $+$ channel of polarimeter $\mathrm{A_0}$, that is, channel $\mathrm{A}_0^+$. It is equal to the probability $P_0(\lambda,\theta_\mathrm{A})$ that this photon exits in the $0$ channel, multiplied by the probability that this same photon, impinging on polarimeter $0$ with an output polarization $\theta_\mathrm{A}$, exits through the output channel $+$:
\begin{equation}\label{pA0plus}
P_\mathrm{0,+}=\cos^2(\lambda+\theta_\mathrm{A})\cos^2(\theta_\mathrm{A}-\varphi_\mathrm{A_0}).
\end{equation}
Similarly, we have
\begin{equation}\label{pA0minus}
P_\mathrm{0,-}=\cos^2(\lambda+\theta_\mathrm{A})\sin^2(\theta_\mathrm{A}-\varphi_\mathrm{A_0}).
\end{equation}

In the general case of a $n$-photon absorption, for a pulse containing exactly $n$ photons all polarized along $\lambda$, the probability that a multiple-photon absorption occurs in either channel of the polarimeter is equal to the probability to that all photons exit in the same output channel $\mathrm{A}_0^+$ plus the probability to that all photons exit in the same output channel $\mathrm{A}_0^-$, that is, $(P_\mathrm{0,+})^n+(P_\mathrm{0,-})^n$. As usual, we have assumed that the photons are independent.

To get the probability of a click in either channel of a polarimeter for our uniformly distributed source of polarized pulses, we just need to average over all possible $\lambda$:
\begin{multline}\label{probpol0}
P_{0,\mathrm{FS}}^{(n)}=\frac{1}{2\pi}\int_0^{2\pi}d\lambda \:
\cos^{2n}(\lambda+\theta_\mathrm{A})\cdot\\
\big(\cos^{2n}(\theta_\mathrm{A}-\varphi_\mathrm{A_0})+\sin^{2n}(\theta_\mathrm{A}-\varphi_\mathrm{A_0})\big),
\end{multline}
and
\begin{multline}\label{probpol1}
P_{1,\mathrm{FS}}^{(n)}=\frac{1}{2\pi}\int_0^{2\pi}d\lambda \:
\sin^{2n}(\lambda+\theta_\mathrm{A})\cdot\\
\big(\cos^{2n}(\theta_\mathrm{A}-\varphi_\mathrm{A_1})+\sin^{2n}(\theta_\mathrm{A}-\varphi_\mathrm{A_1})\big),
\end{multline}

Naturally, in the case of a single-photon absorption, for pulses containing exactly one photon, we get a probability of detection in each polarimeter that is identically equal to $\frac{1}{2}$:
\begin{equation}\label{fssingle}
    P_{0,\mathrm{FS}}^{(1)}=P_{1,\mathrm{FS}}^{(1)}=\frac{1}{2}.
\end{equation}
This illustrates the fact that in this single-photon absorption case the sampling is fair. This is also the result that one would expect from a genuine source of entangled photons: a photon impinging on a polarizing beam-splitter will be detected with probability 1 in either of the two output channels.

However, as soon as we consider a two-photon absorption case, we obtain
\begin{equation}\label{fstwo}
    P_{0,\mathrm{FS}}^{(2)}=P_{1,\mathrm{FS}}^{(2)}=\frac{3}{32} \big(3 + \cos 4(\theta_\mathrm{A}-\varphi_\mathrm{A})\big),
\end{equation}
and similarly in the three-photon absorption case, we get
\begin{equation}\label{fsthree}
    P_{0,\mathrm{FS}}^{(3)}=P_{1,\mathrm{FS}}^{(3)}=\frac{5}{128} \big(5 + 3\cos 4(\theta_\mathrm{A}-\varphi_\mathrm{A})\big),
\end{equation}
where $\varphi_\mathrm{A}$ is either $\varphi_\mathrm{A_0}$ or $\varphi_\mathrm{A_1}$.

\begin{figure}
  \includegraphics[width=9cm]{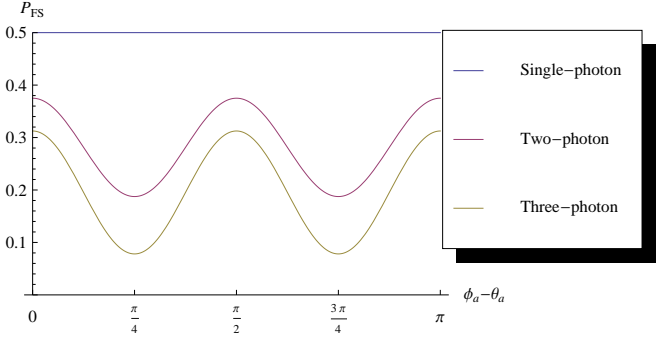}\\
  \caption{Probability of a click in a polarimeter (0 or 1) in the Fair Sampling test as a function of $\theta_\mathrm{A}-\varphi_\mathrm{A}$. In the single-photon case, the probability of click in a polarimeter is independent of the measurement settings: the sampling is fair. As soon however as Eve uses two-photon absorption in order to drive a violation of Bell inequalities in Alice's and Bob's detectors, they can spot the resulting unfairness of the sampling, due to the state-adaptivity of the attack, when $\theta_\mathrm{A}-\varphi_\mathrm{A}$ is varied (which is done automatically if $\varphi_\mathrm{A}$ is fixed, and if $\theta_\mathrm{A}$ is varied randomly from pair to pair as is already the case in an entanglement-based QKD protocol).}\label{fig:FStestresult}
\end{figure}

The average probability to get a click in a polarimeter (0 or 1) becomes clearly dependent on the measurement setting $\theta_\mathrm{A}$ and the polarimeter setting $\varphi_\mathrm{A})$, which betrays the state-adaptivity of the attack, and the resulting unfair sampling (see Fig.\ref{fig:FStestresult}).

Alice can perform this test locally, and unilaterally, in order to spot the unfairness of the sampling due to the state-adaptivity of Eve's attack. She only needs to compare the sum of single counts in the $+$ and $-$ output channel of a polarimeter when $\theta_\mathrm{A}-\varphi_\mathrm{A}$ is varied. She can do so by keeping the settings of her polarimeters fixed, for instance at $\varphi_\mathrm{A}=\varphi_\mathrm{A_0}=\varphi_\mathrm{A_1}=0$, while $\theta_\mathrm{A}$ is randomly switched from pair to pair, as is already the case in both Ekert and BBM92 protocol. Naturally, Bob can perform the test unilaterally and locally as well.

\subsection{Advantages of the Fair Sampling test against other attacks}

Note that the setup used for the Fair Sampling test would also make the other attacks against QKD protocols more complicated (if not impossible) to implement.

The faked-state attack \cite{Makarov07,Makarov10a,Makarov10} consists of Eve impersonating Bob, intercepting the signal that is intended for him and using the exact same procedure that Bob would have implemented. In order to hide her presence, Eve's strategy consists of sending to Bob a signal that will deterministically give him the exact same measurement result as measured by Eve, but only in those cases in which Eve and Bob turned out to be performing their measurements in the same basis. In the remaining cases, when their bases are diagonal to each others, Eve's signal must not trigger any detection in any of Bob's detectors.

To implement this idea, Bob's detectors are blinded to single-photon detection by Eve. After each detection, Eve sends a signal that produces a click in one of Bob's detectors only when it is fully reflected or fully transmitted by its polarizing beamsplitter. So, whenever the bases chosen at random by Eve and Bob are diagonal with respect to each other, the signal sent by Eve is split in half at Bob's polarizing beamsplitter and is insufficient to produce a click in either detector.

This attack would however be immediately visible with our fair sampling test, because it would lead to double-clicks in the polarimeters. Locally, in the cases where Eve performs her measurement in the same basis $\theta_\mathrm{A}$ as Bob, everything would be fine when $|\varphi_\mathrm{A}-\theta_\mathrm{A}|=0$, because the signal sent by Eve would entirely go to one detector only ($A_0^+$ if Eve sent a 0, and $A_1^+$ if she sent a 1). But things would go wrong when $|\varphi_\mathrm{A}-\theta_\mathrm{A}|=\pi/4$ because then the signal sent by Eve would be split evenly inside the corresponding polarimeter ($A_0$ if she sent a 0, or $A_1$ if she sent a 1), so that no click would occur at all. If Eve tries to remedy this situation by increasing the signal, it would only result in both detectors in a polarimeter clicking at the same time (and then it would incidentally also lead to double-clicks when $|\varphi_\mathrm{A}-\theta_\mathrm{A}|=0$). So, the faked-state attack would fail in a fair sampling test setup because it would either lead to double-clicks or to no click at all when $|\varphi_\mathrm{A}-\theta_\mathrm{A}|=\pi/4$.

In the time-shift attack \cite{Lo09a}, when $|\varphi_\mathrm{A}-\theta_\mathrm{A}|=\pi/4$ a photon exiting the PBS in a given channel $i$, would be dispatched randomly and with equal probability to either detector $A^+_i$ or detector $A^-_i$. It would thus become difficult, if not impossible, for Eve to adjust the time-shift associated to a specific polarimeter $i$. She would not have to deal with just one time-dependent detection pattern in each output channel $i$, but with two random ones, depending on whether a photon ends up in $A^+_i$ or in $A^-_i$. Note that this is reminiscent of the idea of randomly switching the bit assignments of the two detectors used in a standard protocol to eliminate the possibility of exploiting the detection efficiency mismatch \cite{Lo09b}.

\section{Conclusion}

Contrary to all existing attacks on QKD protocols, the multiple-photon absorption attack does not require Eve to be physically located between Alice and Bob and to intercept anything that was intended for either of them, as in a typical intercept-and-resend attack \cite{Makarov07,Makarov10a,Makarov10}. The only requirement is that Eve has managed at some point in the past to replace the source of entangled state by her own mixture of separable states. She just needs to know the state $\lambda$ of each pulse and that is something that Eve could have set deterministically in the source. Once this is done, the security of the QKD protocol is compromised by anyone who happens to know the polarization of the pulses as a function of time $\lambda(t)$.

By mimicking the statistics of an entangled state well enough to pass the security checks normally undertaken by Alice and Bob, the attack can in principle work against any entanglement-based quantum key distribution protocol. The attack works in particular for both Ekert and BMM92 protocols, independently of which protocol is chosen by Alice and Bob.

While there already exist explicit attacks on the BBM92 protocol \cite{Zhao08,Makarov10,Makarov10a}, to our knowledge the multiple-photon absorption attack is in fact the only explicit attack against Ekert protocol with light sources.

As long as a device-independent quantum key distribution \cite{Acin07,Pironio09} is beyond technological reach \cite{Lo09a,Lo09b}, it is therefore fundamental for Alice and Bob to either prevent altogether the possibility of multiple-photon absorptions in their detectors, or to include a fair sampling test in the protocol to be able to detect such an attack. The fair sampling test can be performed unilaterally on either side without introducing any loss, and it should also make the task more complicated if not impossible for other existing attacks on QKD protocols.

\begin{acknowledgments}
Irina Basieva is supported by a grant from the Swedish Institute.
\end{acknowledgments}

\end{document}